\documentclass{article}
\usepackage{spconf,amsmath,graphicx}
\usepackage[table,xcdraw]{xcolor}
\usepackage{multirow}
\usepackage{booktabs}
\usepackage{amsfonts}
\usepackage{cite}

\title{Towards Quantifying and Reducing language mismatch effects \\in Cross-lingual Speech Anti-Spoofing}
\name{Tianchi Liu$^{1}$, Ivan Kukanov$^{2}$, Zihan Pan$^{1}$, Qiongqiong Wang$^{1}$, Hardik B. Sailor$^{1}$, Kong Aik Lee$^{3}$}
\address{\fontsize{11pt}{10pt}\selectfont $^{1}$ Institute for Infocomm Research (I$^2$R), Agency for Science, Technology and Research (A$^\star$STAR), Singapore\\
\fontsize{11pt}{10pt}\selectfont $^{2}$ KLASS Engineering and Solutions, Singapore \ \fontsize{11pt}{10pt}\selectfont $^{3}$ The Hong Kong Polytechnic University, Hong Kong}
\begin{document}
%
\maketitle
\begin{abstract}
The effects of language mismatch impact speech anti-spoofing systems, while investigations and quantification of these effects remain limited. Existing anti-spoofing datasets are mainly in English, and the high cost of acquiring multilingual datasets hinders training language-independent models. We initiate this work by evaluating top-performing speech anti-spoofing systems that are trained on English data but tested on other languages, observing notable performance declines. We propose an innovative approach - \textbf{Acc}ent-based data \textbf{e}xpansio\textbf{n} via \textbf{T}TS (\textbf{ACCENT}), which introduces diverse linguistic knowledge to monolingual-trained models, improving their cross-lingual capabilities. We conduct experiments on a large-scale dataset consisting of over 3 million samples, including 1.8 million training samples and nearly 1.2 million testing samples across 12 languages. The language mismatch effects are preliminarily quantified and remarkably reduced over 15\% by applying the proposed ACCENT. This easily implementable method shows promise for multilingual and low-resource language scenarios.
\end{abstract}
\section{Introduction}
\label{sec_intro}
With the rapid advancement of generative models in speech synthesis, synthetic speech is increasingly integrated into daily life and is used in various applications. Speech synthesis is capable of replicating an individual’s voice, emotions, and speaking style, and even generate speech in multiple languages for a given person~\cite{zhang19e_interspeech, 9606610, Nekvinda2020One}. However, this progress also presents significant risks to voice biometrics. Advanced speech generation technologies can be exploited to attack automatic speaker verification (ASV) systems and compromise users~\cite{Evans2013Spoofing, Wu2015Spoofing, chen2024neural, negroni2024analyzing}. 
Finding effective ways to detect spoofing attacks and protect users from spoofed speech is important. This is especially important for the ethical use of generative models.

\begin{figure}[t]
\centerline{\includegraphics[scale=0.052]{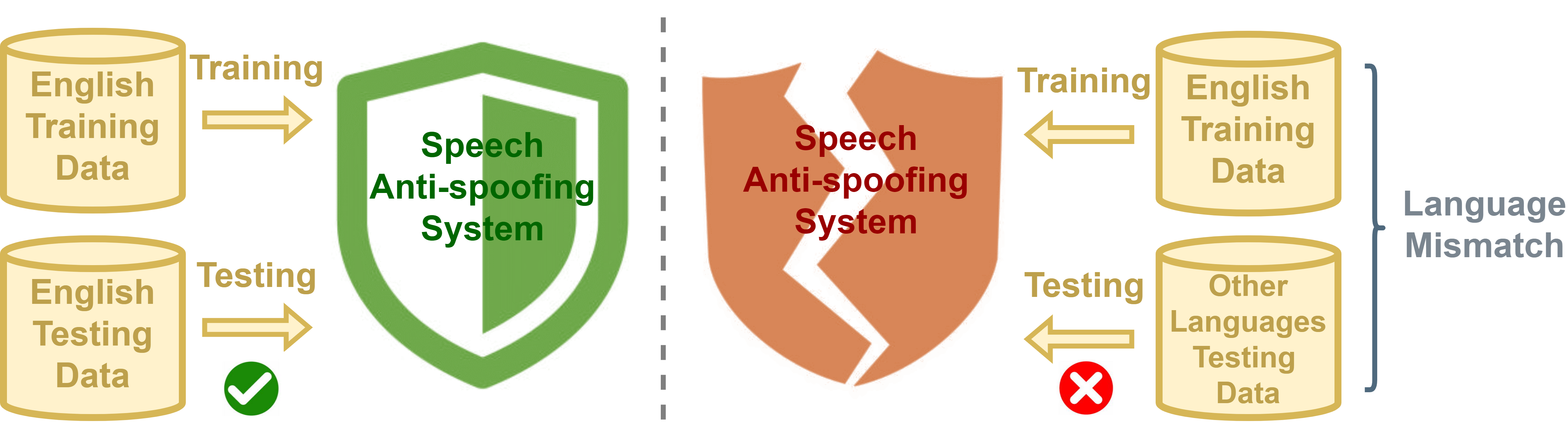}}
\caption{Illustration of language mismatch effects on speech anti-spoofing systems. The English data-trained system works well with English data (left) but fails with other languages (right).}
\label{fig_lang_mis}
\vspace{-0.1 in}
\end{figure}

{As illustrated in Fig.~\ref{fig_lang_mis}, the language mismatch effects refer to the decrease in performance resulting from discrepancies between the languages used during training and inference. These effects have been extensively studied across various speech-related tasks such as ASV, automatic speech recognition (ASR), and speech generation~\cite{zhang19e_interspeech, Misra2014Spoken,Li2017Cross}. 
However, its impact on spoofing detection systems remains underexplored. Our objective is to shed light on this issue.
}

The study on human perception of audio deepfakes demonstrates a positive correlation between language knowledge and the ability of humans to detect audio deepfakes. Native speakers are better at identifying synthetic speech in their language~\cite{Muller2022Human}.
Similar to the findings in human perception studies, a direct approach for the neural system to address language mismatch effects is to gather a substantial amount of high-quality data in different languages for training purposes. However,
existing anti-spoofing datasets are primarily in English~\cite{ASVspoof2015,ASVspoof2017,ASVspoof2019,ASVspoof2021,Idiap_avspoof,khalid2021fakeavceleb}, and acquiring multilingual datasets is costly and challenging, especially for low-resource languages. This situation makes training language-independent models difficult. Consequently, models trained on English data often fail to detect fake speech in other languages.

{Our work seeks to quantify the language mismatch effects and mitigate them in models trained exclusively on English data, aiming to develop robust, language-independent anti-spoofing models.}
{Building on the success of neural vocoders in training anti-spoofing models~\cite{10094779}, we propose an approach to leverage the linguistic knowledge embedded in the Text-to-Speech (TTS) model, which supports multiple languages and accents, to augment an English-only dataset during the vocoding process.} 
{We name this method 
`\textbf{Acc}ent-based data \textbf{e}xpansio\textbf{n} via \textbf{T}TS (\textbf{ACCENT})'.}
{To the best of our knowledge, this paper is the first to investigate and quantify language mismatch effects in speech anti-spoofing.} 

The main contributions are summarized as follows:

\begin{itemize}
\item We explore the language mismatch effects on top-performing CMs. 
\item We propose a novel data expansion method to quantify and reduce language mismatch effects in  speech anti-spoofing models trained on English-only data.
\item We create two multilingual datasets across 10 languages for testing in cross-lingual and low-resource language scenarios.
\item {We are the first to evaluate the effect of language mismatch in the context of  deepfake singing.}
\end{itemize}

\section{Methodology}
\label{sec_accent}

\subsection{Validate the Language Mismatch Effect}
\label{subsec_valideffect}
Studies on {automatic speech recognition} systems show that models trained on multiple languages are more robust against unseen languages compared to those trained only on a single language~\cite{conneau21_interspeech}, highlighting the significance of addressing language mismatch effects. While the language effect has been validated in many speech-related tasks~\cite{Misra2014Spoken,Li2017Cross, conneau21_interspeech}, it hasn't been explored in the field of speech anti-spoofing. 

To fill this gap, we design experiments using an English dataset for training and evaluate the models on test sets both in English and in other languages.
{To ensure that the performance differences are primarily due to language mismatch effects rather than a single model's incidental better adaptation to a specific domain, we re-implemented four state-of-the-art (SOTA) models, including their variants, for a total of seven models:}
SENet~\cite{SEnet} of 34 and 50 layers, SE-Res2Net~\cite{Li2021Replay}, SCG- and MLCG-Res2Net~\cite{li2021channel}, and AASIST/AASIST-L~\cite{Jung2022AASIST}.
We also carefully select the English and other language samples from the same dataset to avoid any biases from different vocoders or data pre-processing differences.
Our findings not only confirm the effect of language mismatch on system performance but also offer an initial quantification. Detailed results and analysis are provided in Section~\ref{sec_discuss}.

\subsection{{Motivations and Hypothesis for ACCENT}}

Obtaining a comprehensive multilingual training set, particularly for low-resource languages, presents a challenge. Moreover, language mismatch has a significant negative impact on performance. To address this, we aim to enhance the cross-lingual capabilities of systems trained on monolingual datasets.

\begin{figure}[t]
\centerline{\includegraphics[scale=0.056]{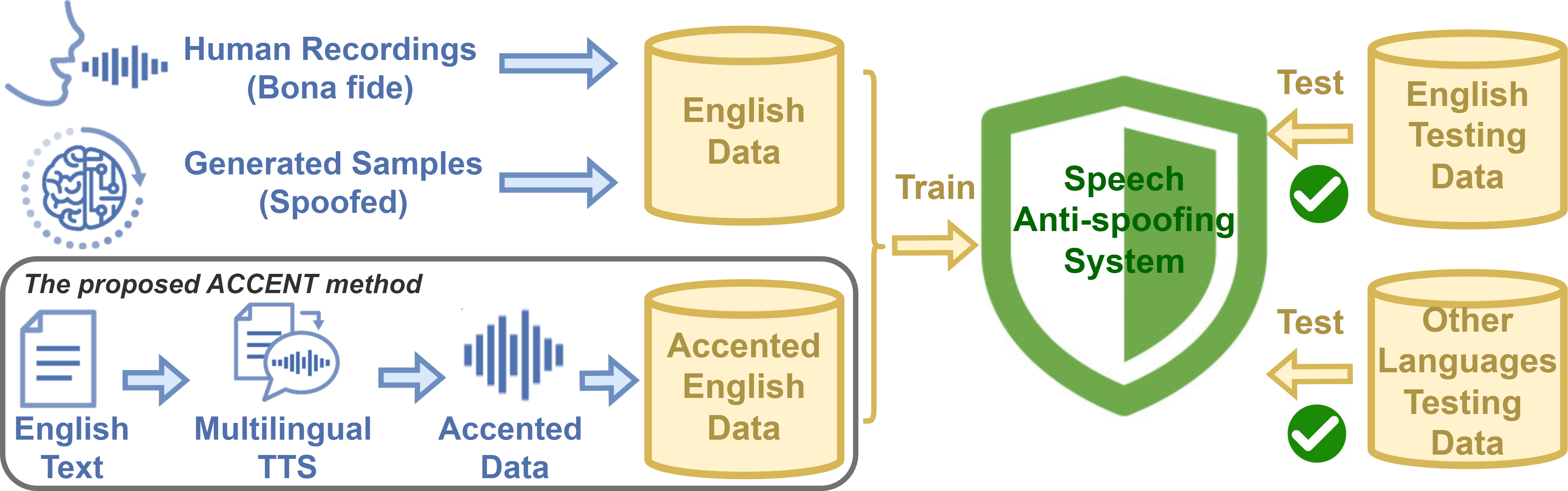}}
\caption{Illustration of the proposed ACCENT method. }
\label{fig_accent_method}
\vspace{-0.15 in}
\end{figure}

The accent refers to distinct pronunciation patterns that mark an individual's speech as belonging to a specific language group~\cite{ARSLAN1996353}. When learning a second language, individuals often retain stylistic features from their native language, blending linguistic characteristics in their speech~\cite{ARSLAN1996353}. Similarly, in samples generated by cross-lingual TTS models, raters may perceive traits resembling those of humans learning to pronounce words in another language through the accent~\cite{zhang19e_interspeech}. We believe that such TTS models with accent will retain the linguistic knowledge from the original language.

Recent work proves the effectiveness of using neural vocoders for training anti-spoofing models rather than building full-fledged speech generation systems~\cite{10094779}.
Building upon this insight, we exploit the diverse linguistic knowledge embedded in TTS models across various languages and accents. 
{By generating accented English data, we enrich the English data with linguistic diversity, which is retained during the vocoding process.}

Our hypothesis posits that utilizing TTS models to generate English data with accents from different languages will retain linguistic knowledge in the vocoding process. This retained information is expected to enhance the model's learning process and improve its capabilities for multilingual and low-resource language scenarios.

Furthermore, the findings in~\cite{10094779} also show that the {anti-spoofing systems trained by vocoders} show promising generalizability, although there's a decline in performance when encountering unseen vocoders. In light of this, we intend to {expand the English-only data set with accents} from as many other languages as possible to enable the models with broad language generalizability, allowing them to effectively handle unseen languages, especially in low-resource language scenarios.
We refer to this novel approach as `\textbf{Acc}ent-based data \textbf{e}xpansio\textbf{n} via \textbf{T}TS (\textbf{ACCENT})'.

\subsection{Methodology of Creating Dataset using ACCENT}
{The pipeline to expand the dataset using the proposed ACCENT method is shown in Fig.~\ref{fig_accent_method}}. Considering that data generated by TTS models are inherently suitable as spoofed data for training anti-spoofing models, and Google TTS (gTTS)\footnote{https://pypi.org/project/gTTS} supports a wide range of languages and accents, we utilize the proposed ACCENT method to create two datasets using gTTS: gTTS-Eng and gTTS-Mix. The {gTTS-Eng} includes 14 English accents and the {gTTS-Mix} is generated by 78 engines of other languages. The creation of these datasets aims to enhance the models' robustness to various English accents and, more crucially, to leverage diverse linguistic knowledge for training language-independent anti-spoofing models.

For {gTTS-Eng}, we randomly use 14 different TTS English accents to voice over English text transcripts chosen from the LibriTTS dataset~\cite{Librispeech}.
For each transcript, we input it into a TTS engine, which then generates a waveform in one of the English accents. Given the relative ease of obtaining English transcripts and ensuring their quality compared to texts from diverse non-English languages,  we develop the gTTS-Mix dataset using 78 different non-English language engines, employing the same method as for gTTS-Eng.
The text transcripts are used to generate speech by one of the 78 engines. Although these utterances are generated using engines for diverse languages, the content remains in English.

{The above data sets are generated using the proposed ACCENT method with gTTS and the LibriTTS dataset. It is important to note that ACCENT is not limited to gTTS and LibriTTS transcripts. Any TTS model with multi-lingual accents and monolingual transcripts can be used to expand the dataset with the proposed ACCENT method.}
\section{Experimental Setup}
\label{sec_exp}

\subsection{Dataset}
\subsubsection{Training Set}

\begin{table}[h]
\vspace{-0.2 in}
\caption{Overview of the training set and its partitions.}
\vspace{0.05 in}

\footnotesize
\centering
\setlength{\tabcolsep}{1.6mm}{
\begin{tabular}{llrrr}
\hline
\toprule
Portion & Dataset  & \# Bona fide & \# Spoof & \# Sum\\ \hline
\midrule
\multirow{6}{*}{I. Public} & ASVSpoof 2015~\cite{ASVspoof2015} & 16,651 & 246,500 & 263,151\\
 & ASVSpoof 2017~\cite{ASVspoof2017} & 3,565 & 0 & 3,565\\
 & ASVSpoof 2019~\cite{ASVspoof2019} & 41,373 & 108,978 & 150,351\\
 & Idiap AVspoof~\cite{Idiap_avspoof} & 15,548 & 55,840 & 71,388\\
 & LibriTTS\cite{Librispeech} & 369,966 & 0 & 369,966\\
 & VCTK~\cite{VCTK} & 44,242 & 0 & 44,242\\ \cline{2-5} 
  & \multicolumn{1}{r}{Subtotal:} & 491,345 & 411,318 & 902,663\\ \hline
\midrule
\multirow{2}{*}{\begin{tabular}[c]{@{}l@{}}II. Accent \\ Aug.\end{tabular}} & gTTS-Eng  & 0 & 48,399 & 48,399\\
 & gTTS-Mix& 0 & \multicolumn{1}{c}{369,674} & 369,674\\ \cline{2-5} 
  & \multicolumn{1}{r}{Subtotal:} & 0 & \multicolumn{1}{l}{418,073} & 418,073\\ \hline
\midrule
 III: Private & Private& 421,135 & 60,504 & 481,639 \\  \hline \midrule
 & \multicolumn{1}{r}{Total:} & 912,480 & 889,895 & 1,802,375\\ \bottomrule
\hline
\label{tab:data_overview}
\end{tabular}
}
\vspace{-0.2 in}
\end{table}

As presented in Table~\ref{tab:data_overview}, we divide the training set into three portions: Portion I primarily consists of public datasets; Portion II contains augmented data created using the proposed ACCENT method; and Portion III comprises private datasets. Both Portions I and II exclusively use English, while in Portion III, 99.93\% of the utterances are in English. The total combined data amounts to over 1.8 million.

\subsubsection{Test Sets}
\label{subsec_testsets}
The test sets in Table~\ref{tab:testset_details} include both monolingual and cross-lingual testing scenarios. For monolingual testing, we include the widely-used ASVspoof series test sets~\cite{ASVspoof2019, ASVspoof2021} and FakeAVCeleb~\cite{khalid2021fakeavceleb}. The WaveFake dataset~\cite{frank2021wavefake}, which encompasses both English and Japanese, is utilized for cross-lingual scenarios. 
The Controlled Singing Voice Deepfake Detection (CtrSVDD) includes samples in both Japanese and Chinese~\cite{zhang2024svdd}. Unlike other test sets that contain speech samples, this test set consists of singing samples. Since the labels of the evaluation set for this dataset has not been released during this work, we use the development set for testing\footnote{https://zenodo.org/records/10467648}\textsuperscript{,}\footnote{https://zenodo.org/records/10742049}.

To simulate cross-lingual and low-resource language scenarios, which are central to this research, we create two new datasets: the \textbf{TTS}-based \textbf{c}ross-\textbf{l}ingual (TTS-CL) dataset and the \textbf{v}oice \textbf{c}onversion-based \textbf{c}ross-\textbf{l}ingual dataset for \textbf{l}ow-resource \textbf{l}anguages (VC-CL3). Altogether, the test sets cover approximately 1.2 million samples across 12 languages. The details of the TTS-CL and VC-CL3 datasets are  as follows:

\begin{table}[t]
\vspace{-0.1 in}
\centering
\caption{Test sets details. Abbr., Lang., and ID indicate abbreviation, language type, and language ID, respectively. {The languages are English (en), Japanese (ja), Chinese (zh), German (de), French (fr), Dutch (nl), Russian (ru), Catalan (ca), Hausa (ha), Indonesian (id), Malay (ms), and Tamil (ta).} }
\vspace{0.05 in}
\footnotesize
\setlength{\tabcolsep}{0.6mm}{
\begin{tabular}{cccccrrr}
\hline
\toprule
 \multirow{2}{*}{} &  \multirow{2}{*}{Dataset} & \multirow{2}{*}{Abbr.} & \multirow{2}{*}{Lang.} & \multirow{2}{*}{ID} & \multicolumn{3}{c}{\# Samples} \\ \cline{6-8} 

\multicolumn{2}{c}{} &  &  & & \multicolumn{1}{c}{\# Bona.} & \multicolumn{1}{c}{\# Spoof} & \multicolumn{1}{c}{\# Sum} \\ \hline
\midrule

\multirow{7}{*}{\rotatebox{90}{Public}} & 2019 LA~\cite{ASVspoof2019} & 19LA & \multirow{5}{*}{\rotatebox{90}{Monolingual}} & en & 7,355 & 63,882 & 71,237 \\
  & 2021 LA~\cite{ASVspoof2021} & 21LA &   & en & 14,816 & 133,360 & 148,176 \\
  & 2021 DF~\cite{ASVspoof2021} & 21DF &  & en & 14,869 & 519,059 & 533,928 \\
  
  & FakeAVCeleb~\cite{khalid2021fakeavceleb} & FAVC &  & en & 10,209 & 11,335 & 21,544 \\
  \cline{5-8} 
  &  &  &  & \multicolumn{1}{r}{} & 47,249 & 727,636 & 774,885\\ 
  
  \cline{2-8}
 &  CtrSVDD Dev~\cite{zhang2024svdd} & -  & \multirow{3}{*}{\begin{tabular}[c]{@{}l@{}}Cross- \\ lingual\end{tabular}}  & ja, zh & 4,650 & 37,078 & 41,728 \\
& \multirow{3}{*}{WaveFake~\cite{frank2021wavefake}} & \multirow{3}{*}{-} & & en & 13,100 & 107,983 & 121,083 \\

   &  & &  &  ja  & 5,000 & 10,000 & 15,000 \\ 
\cline{5-8} 
 &  &  &  & \multicolumn{1}{r}{}  & 22,750 & 155,061 & 177,811\\
   \hline
   
\midrule
\multirow{12}{*}{\rotatebox{90}{Ours}} &  \multirow{6}{*}{\begin{tabular}[c]{@{}l@{}}\textbf{TTS}-based \\ \textbf{C}ross-\textbf{l}ingual\end{tabular}}
& \multirow{6}{*}{\begin{tabular}[c]{@{}l@{}}TTS \\ -CL\end{tabular}} & \multirow{6}{*}{\begin{tabular}[c]{@{}l@{}}Cross- \\ lingual\end{tabular}} & de & 7,427 & 37,135 & 44,562 \\
 &  &  & &  fr  & 8,649 & 43,245 & 51,894 \\
 &  &  & &  nl  & 6,494 & 30,725 & 37,219 \\
 &  &  & &  ru  & 9,599 & 47,990 & 57,589 \\
 &  &  & & zh  & 2,971 & 14,855 & 17,826 \\ \cline{5-8} 
  &  &  &  & \multicolumn{1}{r}{} & 35,140 & 173,950 & 209,090\\ 
  
 \cline{2-8}
 & \multirow{6}{*}{\begin{tabular}[c]{@{}c@{}}\textbf{VC}-based \\ \textbf{C}ross-\textbf{l}ingual \\ dataset for\\ \textbf{L}ow-resource \\ \textbf{L}anguages \end{tabular}} & \multirow{6}{*}{\begin{tabular}[c]{@{}l@{}}VC- \\ CL3\end{tabular}}  & \multirow{6}{*}{\begin{tabular}[c]{@{}l@{}}Cross- \\ lingual\end{tabular}}  & ca & 3,644 & 3,644 & 7,288 \\
 &  &  & & ha  & 4,176 & 4,176 & 8,352 \\
 & & &  & id   & 3,616 & 3,616 & 7,232 \\
 &  & & & ms   & 3,740 & 3,740 & 7,480 \\
 & & &  & ta   & 3,335 & 3,335 & 6,670 \\ \cline{5-8} 
  &  &  &  & \multicolumn{1}{r}{} & 18,511 & 18,511 & 37,022\\ 

     \hline
   
\midrule
 &  &  &  \multicolumn{2}{r}{Total:} & 123,650 & 1,075,158 & 1,198,808\\ 
\bottomrule
\hline
\label{tab:testset_details}
\end{tabular}
}
\vspace{-0.3 in}
\end{table}

\textbf{TTS-CL} dataset is generated using the CSS10~\cite{park2019css10} subset of five languages: German, French, Dutch, Russian, and Chinese, serving as {bona fide} speech samples. CSS10 dataset is a single native speaker dataset. Synthetic speech is generated with cross-language TTS approach \cite{Nekvinda2020One}. 
{The cross-language TTS approach is based on a Tacotron-style model~\cite{Shen2018Natural} that outputs a generated spectrogram, which is then passed to the WaveRNN vocoder~\cite{Kalchbrenner2018Efficient} to generate a waveform.}
The vocoder is trained on the CSS10~\cite{park2019css10} dataset.

\textbf{VC-CL3} dataset is constructed to investigate the efficacy of the proposed method in low-resource language scenarios. The dataset encompasses five languages: Catalan, Hausa, Indonesian, Malay, and Tamil.
We build this dataset with the voice conversion model available from the Coqui-TTS toolkit\footnote{https://github.com/coqui-ai/TTS}. Bona fide data are sourced from the FLEURS dataset~\cite{FLEURS}. The target speaker for voice conversion is randomly selected from the dataset of the same language. In order to maintain a balance between bona fide and spoofed speech, each bona fide speech is spoofed only once.

\subsection{Training Strategy}

We use Adam optimizer with a weight decay of $1 \times 10^{-4}$ and a learning rate scheduler with $1000$ warm-up steps. 
The learning rate decrease proportionally to the inverse square root of the step number. The negative log-likelihood loss is used. Training ceases if there is no improvement in validation performance for 12 epochs. The best model on the validation set is used for testing. Three type of data augmentations\footnote{https://github.com/iver56/audiomentations} of adding Gaussian noise, pitch shifting, and time stretching are applied during training. We train all the models using the same setup, with the exception of two specific cases: (1). The models in Table~\ref{tab:baseline_crosscorpus} are trained with a patience of 15 epochs.
(2). The self-supervised learning (SSL)-based systems are trained using WavLM large~\cite{9814838} coupled with a Long Short-Term Memory (LSTM) with hidden dimension of 192. The training starts with 5-epoch warm up, then exponentially decaying. The SSL model is fixed for the first 10 epochs.

\begin{table}[h]
\vspace{-0.15 in}
\footnotesize
\centering
\caption{Dataset splits for training, validation, and evaluation. {`Eng', `Mix', and `Ds' refer to gTTS-Eng, gTTS-Mix, and a combination of the two, respectively. These three data split configurations have been downsampled to align with the training and validation set sizes used in Portion I. `Full' refers to the unaltered set of both gTTS-Eng and gTTS-Mix datasets.}}
\vspace{0.1 in}
\setlength{\tabcolsep}{2mm}{
\begin{tabular}{lrrccr}
\hline
\toprule
Portion & \multicolumn{1}{c}{Train} & \multicolumn{1}{c}{Valid} & \multicolumn{1}{c}{Ratio} & \multicolumn{1}{c}{Eval} & \multicolumn{1}{c}{Total} \\ \hline
\midrule
I & 510,416 & 127,605 & 4:1 & \multirow{7}{*}{264,642} & 902,663 \\
{I+II (Eng)} & 510,416 & 127,605 & 4:1 &  & 902,663 \\
{I+II (Mix)} & 510,416 & 127,605 & 4:1 &  & 902,663 \\
I+II (Ds) & 510,416 & 127,605 & 4:1 &  & 902,663 \\
I+II (Full) & 844,874 & 211,220 & 4:1 &  & 1,320,736 \\

I+III & 895,728 & 223,932 & 4:1 &  & 1,384,302 \\
I+II+III & 1,230,186 & 307,547 & 4:1 &  & 1,802,375\\
\bottomrule
\hline
\end{tabular}
}
\label{tab:train_split}
\vspace{-0.1 in}
\end{table}

The data in Table~\ref{tab:train_split} are divided into training and validation sets in a 4:1 ratio. Evaluation sets from the public portion are not included in either training or validation. The three portions outlined in Table~\ref{tab:data_overview} are divided into seven configurations to assess the efficacy of the proposed ACCENT method.

\section{Results and Discussion}
\label{sec_discuss}

\subsection{Effects of Language Mismatch on SOTA Models}

As detailed in Section~\ref{subsec_valideffect}, we train seven SOTA models using the ASVspoof2019 LA~\cite{ASVspoof2019} dataset in an English-only condition, and test on English datasets of 19 LA~\cite{ASVspoof2019}, 21 LA~\cite{ASVspoof2021}, 21 DF~\cite{ASVspoof2021}, FakeAVCeleb~\cite{khalid2021fakeavceleb}, and a cross-lingual dataset, WaveFake~\cite{frank2021wavefake}. The results are reported in Table~\ref{tab:baseline_crosscorpus}. 


\begin{table}[t]
\centering
\footnotesize
\vspace{-0.08 in}
\caption{Cross-corpus performance of top-performing systems trained on ASVspoof 2019 LA in EER (\%). `↑' shows the relative increase when the models are tested on samples that include Japanese (WF$_{m}$) compared to those tested only in English (WF$_{e}$). Para. is the number of parameters in million. FAVC indicates FakeAVCeleb dataset.}
\vspace{0.05 in}
\label{tab:baseline_crosscorpus}
\setlength{\tabcolsep}{0.9 mm}{
\begin{tabular}{ccccccccc}
\hline
\toprule

 \multirow{2}{*}{Model} & \multirow{2}{*}{Para.}  & \multicolumn{4}{c}{English Test Sets} & \multicolumn{3}{c}{WaveFake}  \\ \cmidrule(r){3-6} \cmidrule(r){7-9}

& & 19LA & 21LA & 21DF & FAVC  &  WF$_{e}$  &  WF$_{m}$ & ↑ (\%)\\
\hline
\midrule
SENet-34 & 1.34  & 8.08 & 19.43 & 22.57  &24.46& \textbf{30.31}& \textbf{39.03} & 28.8 \\

SENet-50 & 1.09  & 10.82 & 20.23 & 23.58   & \textbf{19.03}& 36.23& 44.57& 23.0 \\

SE-Res2Net &  0.85  &	2.89 & 20.16 & 19.35 & 21.86 & 34.66 &39.45 &13.8\\

SCG-Res2Net &  0.95  &	2.81 & 15.77 & \textbf{18.47} & 26.97  & 30.66 &  39.16 & 27.7 \\

MLCG-Res2Net & 0.94  & 2.80 & 23.59 & 18.63& 28.57 &  37.36 &  41.33 & 10.6\\

AASIST-L & \textbf{0.09} &	0.98 & 12.69 & 20.05 &42.33 & 40.86 & 49.53 &21.2 \\

AASIST & 0.30  & \textbf{0.83} & \textbf{11.46} & 21.07&43.83  &46.19& 53.36& 15.5 \\
\bottomrule
\hline
\end{tabular}}
\vspace{-0.15 in}
\end{table}

We observe that, on the same WavFake dataset, simply changing the testing language from English (\textbf{WF$_{e}$)} to a mix of Japanese and English (\textbf{WF$_{m}$}) results in a 20.1\% increase in the average Equal Error Rate (EER) across seven models. 
It is important to note that, according to~\cite{frank2021wavefake}, which provides information on English and Japanese fake samples, the TTS models used in WF$_e$ and WF$_m$ are comparable. The only difference is that WF$_m$ includes Multi-band MelGAN, which uses a larger generator version than the MelGAN generator present in both datasets~\cite{frank2021wavefake}. Since the generative models in both test sets are largely similar, we attribute the performance differences primarily to language mismatch effects.
{The re-implementation of seven SOTA models ensures that performance differences are attributed to language mismatch effects, rather than any single model’s incidental advantage in adapting to a specific domain.} This experiment validates the language mismatch effect and provides an initial quantification of this effect between English and Japanese.
Furthermore, SCG-ResNet is selected as our primary model for following experiments due to its generally superior performance.

\subsection{Evaluation for the proposed ACCENT method}
\label{subsec_eval_accent}
\begin{table*}[h]
\vspace{-0.05 in}
\footnotesize
\centering
\caption{The performance in EER (\%) of systems with or without the proposed ACCENT method. Aug. signifies whether the system is trained with data augmentations.  `Avg. ↓' signifies the average relative reduction in EER. The results, presented in the format of mean ${\pm}$ standard deviation are conducted twice. Models with our proposed ACCENT are highlighted in \colorbox{gray!20}{grey}.}
\vspace{0.01 in}
\setlength{\tabcolsep}{2mm}{
\renewcommand{\arraystretch}{1.02}{
\begin{tabular}{cclrcllllll}
\hline
\toprule
\multirow{2}{*}{\#} & \multirow{2}{*}{Model} & \multirow{2}{*}{Portion} & \multirow{2}{*}{\#Sample} & \multirow{2}{*}{Aug.} & \multicolumn{3}{c}{Cross-lingual Test Set} & \multicolumn{3}{c}{English Test Set} \\
\cmidrule(r){6-8} \cmidrule(r){9-11}
 &  &  &  &  & VC-CL3 & TTS-CL & Avg. ↓ & 2019 LA & 2021 LA & 2021 DF \\
 \hline
\midrule
1 & \multirow{5}{*}{\begin{tabular}[c]{@{}c@{}}SCG-\\ Res2Net \\ \cite{li2021channel}\end{tabular}} & I & 510,416  &$\checkmark$& 33.54 $_{\pm 3.91}$ & 21.09 $_{\pm 4.34}$ & Benchmark  & \textbf{3.47 $_{\pm 0.45}$} & 12.28 $_{\pm 0.49}$ & 25.72 $_{\pm 1.91}$    \\

2 &  & \cellcolor[HTML]{EFEFEF}I+II (Eng) & 510,416 & $\checkmark$ & 34.25 & 20.01 & -1.5\%  & 3.60  &  \ \ \textbf{8.91} & \textbf{20.93}  \\
3 &  & \cellcolor[HTML]{EFEFEF}I+II (Mix) & 510,416 & $\checkmark$ & 26.72 & 18.82 & -15.6\% & 4.77 & 11.42 & 27.03  \\
4 &  & \cellcolor[HTML]{EFEFEF}I+II (Ds) & 510,416 & $\checkmark$ & \textbf{25.34 $_{\pm 3.71}$ } & \textbf{17.67 $_{\pm 2.19}$ } & \textbf{-20.3\%}  & 4.58 $_{\pm 0.43}$ & 13.53 $_{\pm 2.58}$ & 25.87 $_{\pm 3.48}$  \\
5 &  & \cellcolor[HTML]{EFEFEF}I+II (Full) & 844,874 & $\checkmark$ & 25.67 $_{\pm 2.52}$ & 17.78 $_{\pm 0.97}$  & -19.6\% & 4.74 $_{\pm 1.39}$ & {11.75 $_{\pm 2.11}$} & {22.94 $_{\pm 2.02}$}   \\  

\hline
 
6 & \multirow{2}{*}{\begin{tabular}[c]{@{}c@{}}\textit{Gemini}- \\ \ Res2Net~\cite{liu2023golden} \ \end{tabular}} & I  & 510,416 & $\checkmark$ & 35.68 & 22.10 & Benchmark   &   \textbf{2.39} & \textbf{11.93} & 19.79  \\
7 &  & \cellcolor[HTML]{EFEFEF}I+II (Ds)  & 510,416 & $\checkmark$ & \textbf{25.21} & \textbf{15.99} & -28.5\%  & 3.92 & 14.25 & \textbf{19.08} \\

\hline

8 & \multirow{2}{*}{\begin{tabular}[c]{@{}c@{}}WavLM- \\ \ Large~\cite{9814838} \ \end{tabular}} & I &  510,416 & \textbf{$\times$}   & \textbf{39.16} & 48.12 & Benchmark   &  \textbf{0.68} &  \ \ 4.96 &  \ \ 5.65  \\
9 &  & \cellcolor[HTML]{EFEFEF}I+II (Ds)  & 510,416 &  \textbf{$\times$}  & 48.04  & \textbf{31.50} & -5.9\% & 0.84 &  \ \ \textbf{3.59} &  \ \ \textbf{4.13} \\

\hline

\midrule
10 & \multirow{2}{*}{\begin{tabular}[c]{@{}c@{}}SCG-\\ Res2Net~\cite{li2021channel}\end{tabular}}  & I+III & 895,728 & \textbf{$\checkmark$} &  5.70 &  1.21 & Benchmark  & 2.99 & \ \   9.42 & 20.55   \\
11 &  & \cellcolor[HTML]{EFEFEF}I+II+III  & 1,230,186 & \textbf{$\checkmark$} & \textbf{4.59} & \textbf{1.04} & -16.7\% & \textbf{2.20} & \ \ \textbf{8.59} & \textbf{17.73} \\ \hline
12 & \multirow{2}{*}{\begin{tabular}[c]{@{}c@{}}\textit{Gemini-} \\ \ Res2Net~\cite{liu2023golden} \ \end{tabular}} & I+III & 895,728  & \textbf{$\checkmark$} & 4.73 & 0.90 & Benchmark   & 1.96 & \ \  \textbf{8.24} & 14.54 \\
13 &  & \cellcolor[HTML]{EFEFEF}I+II+III & 1,230,186 & \textbf{$\checkmark$} & \textbf{3.95} &   \textbf{0.83} & -12.1\% & \textbf{1.74} & 10.12 & \textbf{13.78} \\

\bottomrule
\hline
\end{tabular}
}
}
\label{tab:accentaug}
\vspace{-0.15 in}
\end{table*}

\begin{figure}[h]
\centerline{\includegraphics[scale=0.126]{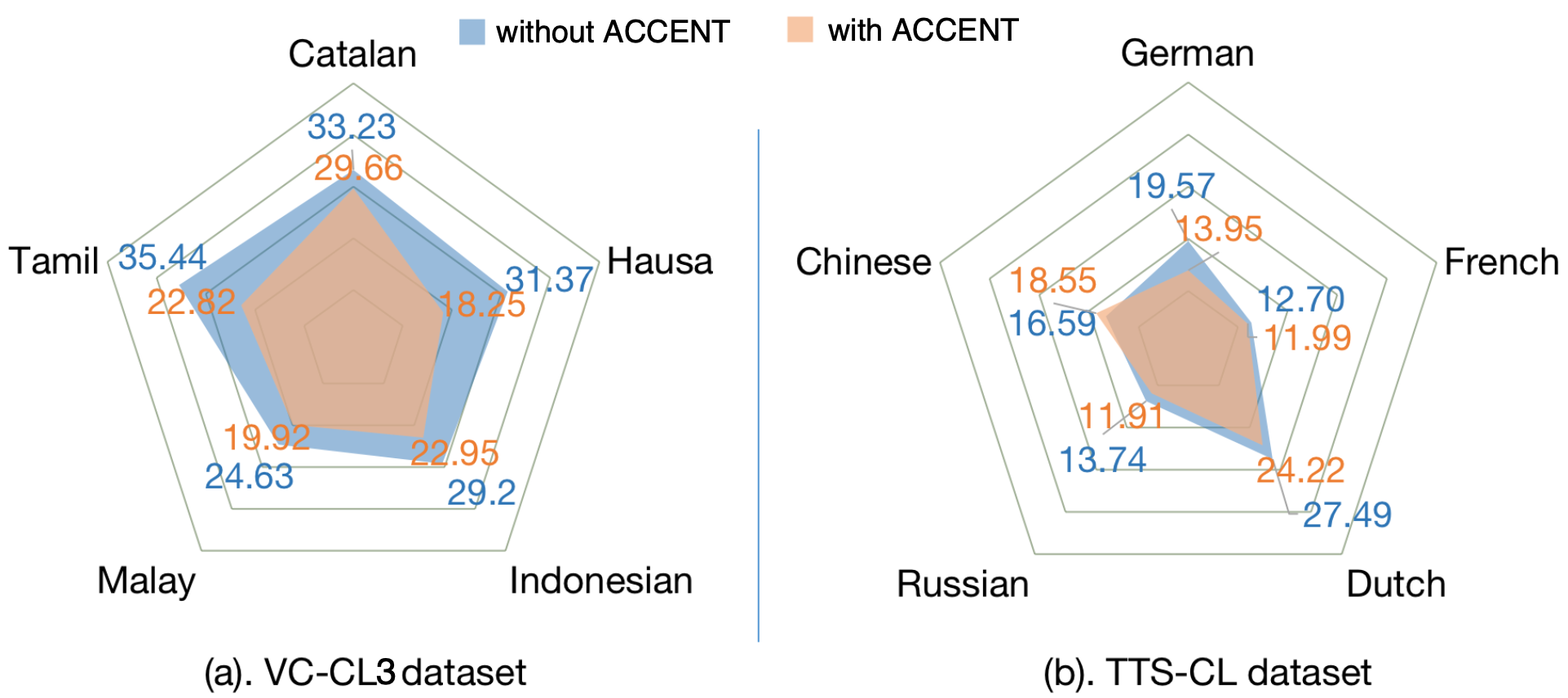}}
\vspace{-0.05 in}
\caption{The radar charts show the EER (\%) performance of systems across ten languages in the VC-CL3 and TTS-CL datasets. Systems without and with the proposed ACCENT method correspond to system 1 and 4 in Table~\ref{tab:accentaug}, respectively. Each system is trained twice, with the better performance displayed in the radar charts.}
\label{fig_visual_low_rees}
\vspace{-0.15 in}
\end{figure}

We compare the performance of the SCG-Res2Net model on two cross-lingual test sets, both with and without the implementation of the proposed ACCENT method. The results are presented in Table~\ref{tab:accentaug}. In this context, Portion I is the public dataset, while Portion II consists of data augmented with our proposed ACCENT method; both are English datasets. A comparison between System 1 and System 5 reveals that the proposed ACCENT method significantly enhances the model's performance on cross-lingual test sets, achieving an average relative improvement of 19.6\%, even when the training data is still exclusively monolingual in English.

\textbf{Excluding the Impact of Data Volume Changes}.
By merging Portions I and II, the total number of training samples increased from 510,416 to 844,874, referred to as \mbox{`I+II (Full)'}. To eliminate the possibility that the performance improvement is due to the increased number of training samples, we downsampled the `I+II (Full)' to match the Portion I, naming this subset `I+II (Ds)'. The training results on this subset, shown as System 4 in Table~\ref{tab:accentaug}, is similar to System 5.  This supports the conclusion that  the improved cross-lingual test outcomes are due to the proposed ACCENT method.

\textbf{Evaluation for Different Languages}. In Figure~\ref{fig_visual_low_rees}, we conduct a detailed investigation of the performance across various languages. The radar charts show that the proposed ACCENT method significantly enhances performance in 9 out of 10 different languages, with the performance in the remaining language, Chinese, also being very close. This demonstrates the robustness of the proposed ACCENT method, which encompasses 78 language accents and 14 English accents, in improving performance across diverse languages.
{Comparing the languages supported by gTTS, most languages in the VC-CL3 and TTS-CL datasets are covered, except for Hausa. Despite this, significant performance gains were observed even for this unsupported language, highlighting the robustness and generalization capabilities of our proposed method in handling unseen languages.}

\textbf{Evaluation for Different Models.} We perform experiments to evaluate the compatibility of the proposed ACCENT method with two other models. 

The \textit{Gemini} model, as presented in~\cite{liu2023golden, liu2024disentangling}, is designed to investigate the effects of temporal and frequency resolutions on speech tasks in the context of 2D-CNN models. The Res2Net structure, known for its ability to extract multi-scale features, has had its performance validated in related speech tasks~\cite{10448049, chen23o_interspeech, chen2024eres2netv2, 9747021}. Therefore, we include \textit{Gemini} Res2Net in our experiments.
According to the results presented in Table~\ref{tab:accentaug}, our proposed ACCENT method shows great compatibility with the \textit{Gemini} Res2Net model, achieving an average of 28.5\% relative improvement. 

The SSL models have been widely explored in anti-spoofing tasks, showing exceptional performance~\cite{tak2022automatic, 10003971, liu2024neural, pan2024attentive}.
In Table~\ref{tab:accentaug}, we investigate WavLM large~\cite{9814838}-based model and observe its suboptimal performance on cross-lingual test sets comparing to conventional models.
Given its superior performance on ASVspoof test sets, and the absence of data augmentation techniques, we believe that the systems 8 and 9 encounter overfitting issues. This could also explain why the TTS-based ACCENT method only leads to performance improvements on the TTS-based dataset, TTS-CL, whereas there is a decrease in performance on the VC-based test set, VC-CL3. We recommend integrating the proposed ACCENT method with augmentations to mitigate overfitting concerns. Further investigation into SSL models will be valuable.

\textbf{Evaluation on English Test Sets}. Our previous experiments clearly show the benefits of the proposed ACCENT method in cross-lingual testing scenarios. We now turn our attention to its effects on the training language, English. The performance outcomes of all systems on the ASVspoof test sets are detailed in the right part of Table~\ref{tab:accentaug}. In more than half of cases, an improvement in performance is observed. These results indicate that while the proposed ACCENT method enhances cross-lingual test performance, it does not negatively impact performance on English test sets.

{\textbf{Ablation Study.} When various language accents are applied, system 3 demonstrates obvious
improvements in multilingual performance over the baseline system 1. This highlights the
effectiveness of the proposed ACCENT method. With augmentation limited to English accents,
system 2 shows a slight advantage over all other systems in English test sets, which is
expected given its English-centric design. Moreover, the combination of both achieves a better balance.}

\textbf{Compatibility with Massive Types of Synthetic Data}. 
{Considering that the proposed method introduces additional TTS models into the data, the observed performance improvement might be attributed to the extra model rather than the linguistic information related to the accent, which enhances the model's robustness.  To minimize this possibility, we included data from Portion III. Portion III contains data generated by over 70 different synthetic models. By introducing a large number of different models, we can minimize the benefit gained from additional TTS models. Additionally, we investigate whether the proposed ACCENT method method remains effective for data generated by a wide variety of synthetic models.} The experimental results from Table~\ref{tab:accentaug} indicate that by enriching the types of synthetic data, there is a significant improvement in cross-lingual performance. Moreover, the concurrent use of the proposed ACCENT method can further improve the performance by 16.7\% (system 10 vs. 11)  and 12.1\% (system 12 vs. 13).

\subsection{Evaluation on Synthetic  Singing  Test Set}

\begin{table}[h]
\vspace{-0.25 in}
\footnotesize
\centering
\caption{Performance in EER (\%) on CtrSVDD Dev. set.   Models with our ACCENT method are highlighted in \colorbox{gray!20}{grey}. `Avg. ↓' signifies the average relative reduction in EER. 
}
\vspace{0.05 in}
\setlength{\tabcolsep}{2mm}{
\begin{tabular}{ccllc}
\hline
\toprule
 \multirow{1}{*}{\#} & \multirow{1}{*}{Model} & \multirow{1}{*}{Portion}  & \multicolumn{1}{c}{CtrSVDD Dev.} & \multicolumn{1}{c}{Avg. ↓} \\ \hline
\midrule
1 & \multirow{4}{*}{\begin{tabular}[c]{@{}c@{}}SCG-\\ Res2Net~\cite{li2021channel}\end{tabular}} & I & {15.85 $_{\pm 1.87}$} & -    \\
4 &  & \cellcolor[HTML]{EFEFEF}I+II (Ds) & \textbf{13.50 $_{\pm 2.22}$} & -14.8\% \\
\cline{3-5}
10 &  & I+III &  \ \ 9.42 & -\\
11 &  & \cellcolor[HTML]{EFEFEF}I+II+III &  \ \ \textbf{8.59 } & -8.8\%  \\
\hline
6 & \multirow{4}{*}{\begin{tabular}[c]{@{}c@{}}\textit{Gemini}- \\ \ Res2Net~\cite{liu2023golden} \ \end{tabular}} & I &   19.18 & -\\
7 &  & \cellcolor[HTML]{EFEFEF}I+II (Ds)  & \textbf{16.17} &  -15.7\% \\
\cline{3-5}
12 &  & I+III &  \ \ \textbf{9.85} & - \\
13 &  & \cellcolor[HTML]{EFEFEF}I+II+III & 10.13 & +2.8\% \\

\bottomrule
\hline
\end{tabular}
}
\label{tab:svdd}
\vspace{-0.10 in}
\end{table}

With the advancements in research on synthetic singing, the detection of synthetic singing has recently garnered more attention in the research community~\cite{zhang2024svdd, guragain2024speech, 10448184, 10446271}. As introduced in Section~\ref{subsec_testsets}, CtrSVDD is a dataset comprising both Japaneses and Chinese, meeting our requirements for cross-lingual performance evaluation. Therefore, we also evaluate the proposed method to determine if it remains effective in the singing voice deepfake detection scenario. The experimental results, reported in Table~\ref{tab:svdd}, indicate that in most cases, the proposed method improves performance. Furthermore, we observe that using data from more diverse generative models in Portion III under the singing condition significantly enhances the model's robustness and improves performance. This finding is consistent with that in Section~\ref{subsec_eval_accent}.

\subsection{Quantifying Language Mismatch Effects}
Currently, the extent to which language mismatch affects the performance of anti-spoofing systems remains unknown.
Based on observations from all of our experiments, {the overall performance improvement with the proposed method is approximately 15\%. Given that not all linguistic information can be preserved during the vocoding process and that} our method cannot perfectly solve the language mismatch effect, we estimate that the language mismatch effects could relatively decrease performance by over 15\%. However, this can vary depending on factors such as the diversity of training data, model architectures, training strategies, and the languages used for testing.

\section{Conclusion}
We initiate our work by investigating the language mismatch effects on several top-performing anti-spoofing systems. 
We observe that their effectiveness decreases by an average of 20\% when tested on a mix of Japanese and English samples, versus only English samples. This confirms the existence of language mismatch effects between English and Japanese.
To broaden this analysis to more languages and enhance cross-lingual capabilities in anti-spoofing models, we introduce a data expansion method named ACCENT. This method integrates diverse linguistic information into monolingual-trained systems through a vocoding process.
The results show that our approach remarkably improves cross-lingual performance by 15\%, without compromising the system's performance in its original training language. This method is promising for multilingual and low-resource language scenarios. Moreover, it holds potential for application in other speech tasks to improve language generalization ability and robustness. Additionally, we are the first to quantify the effects of language mismatch on performance in speech anti-spoofing, offering a preliminary estimate of over 15\% relative reduction.

\section{Acknowledgements}
This work is supported
by the National Research Foundation, Prime Minister’s Office, Singapore, and the Ministry of Digital Development and Information, under its Online Trust and Safety (OTS) Research Programme (MCI-OTS-001). 
The opinions in this material are those of the authors and do not reflect the views of the aforementioned institutions.
\bibliographystyle{IEEEbib_limit6}
\bibliography{strings}

\end{document}